\documentstyle[prd,aps,epsfig]{revtex}

\begin{document}
\draft

\title{Exotic mesons, locking states and their role in
the formation of the confinement barrier }
\author{V. V. Anisovich\thanks{anisovic@thd.pnpi.spb.ru}}
\address{St.Petersburg Nuclear Physics Institute, Gatchina, 188350,
Russia}
\author{D. V. Bugg\thanks{bugg@v2.rl.ac.uk}}
\address{Queen Mary and Westfield College, London E14NS, UK}
\author{A. V. Sarantsev
\thanks{vsv@hep486.pnpi.spb.ru and andsar@v2.rl.ac.uk}}
\address{St.Petersburg Nuclear Physics Institute, Gatchina, 188350,
Russia}
\maketitle

\begin{abstract}
We discuss a mechanism by which a set of $q\bar q$-states mixes with
an exotic one, so that the exotic state accumulates the widths of the
overlapping $\bar qq$ resonances. The broad state created this way
acts as a 'locking' state for the others. Using results of a previous
analysis of the $(IJ^{PC}=00^{++})$-wave, we estimate the mean radius
squared of the broad state $f_0 (1530^{+90}_{-250})$; it appears to be
distinctly larger than the mean squared radii of the locked narrow
states. This supports an idea about the constructive role of broad
states in forming the confinement barrier.
\end{abstract}


\twocolumn

\narrowtext
The objective of this letter is to introduce a qualitatively new idea
having two components. The idea is that one resonance which overlaps
other resonances may become broad via mixing and may in turn reduce
the widths of the others; it thus acts as a confinement barrier for
narrow states. We call the broad state a `locking' state. We draw
attention to experimental evidence for this idea and suggest that
formation of broad states may be a general phenomenon involving
exotic states.

{\bf Effect of accumulation of widths in the $K$-matrix
approach.}
To examine the mixing of non-stable states in a pure form, let us
consider an example with three resonances decaying into the same
channel. In the $K$-matrix approach, the amplitude we consider reads:
\begin{equation}
A = K(1-i\rho K)^{-1}\ ,\;
K = g^2\sum_{a=1,2,3}(M^2_a-s)^{-1}\ .
\end{equation}
Here, for purposes of illustration, we take $g^2$ to be the same for
all three resonances, and make the approximations that: (i)~the phase
space factor $\rho$ is constant, and (ii)~$M^2_1=m^2-\delta$,
$M^2_2=m^2$, $M^2_3=m^2+\delta$. Fig.\ 1 shows the location of poles
in the complex-$M$ plane $(M=\sqrt s)$ as the coupling $g$ increases.
At large $g$, which corresponds to a strong overlapping of the
resonances, one resonance accumulates the widths of the others while
two counterparts of the broad state become nearly stable.

{\bf Resonance structure in the $(IJ^{PC}=00^{++})$-wave. }
This simple model may be compared with what is known of the actual
situation for the $\pi \pi $ S-wave. The analysis of this wave has
been pursued in a series of papers using a variety of technical
approaches: the $T$-matrix \cite{bsz} and multichannel $K$-matrix
approaches \cite{km1900} and the dispersion relation $N/D$-method
\cite{n/d}. The most recent $K$-matrix analysis was done using data
from GAMS \cite{gams}, the Crystal Barrel collaboration \cite{cbc} and
the BNL group \cite{bnl}. It allows us to determine resonance
structure in this wave in the mass range up to 1900 MeV. There are
five states in this mass region. Four of them may be identified as
$q\bar q$ states, members of $1^3 P_0 q\bar q$ and $2^3 P_0 q\bar q$
nonets; the fifth state in the mass region 1250 - 1650 MeV is an extra
one not accomodated by $q\bar q$ systematics. The restored coupling
constants for decays into channels $\pi\pi$, $K\bar K$, $\eta\eta$ and
$\eta\eta '$ show that the extra state has at least a 40\% - 50\%
admixture from the lightest scalar glueball.

Parameters found in \cite{km1900} for the $K$-matrix $0^{++}$
amplitude allow a direct comparison of the mixing dynamics with the
simple model of Fig.\ 1. We introduce a scaling parameter $\xi $ into
the $K$-matrix elements,
\begin{equation}
K_{ab}=\sum_\alpha \frac{g^{(\alpha)}_a g^{(\alpha)}_b}{
M^2_\alpha-s}+f_{ab} \to \sum\xi
\frac{g^{(\alpha)}_ag^{(\alpha)}_b}{M^2_\alpha-s}+\xi f_{ab}\ ,
\end{equation}
and vary $\xi$ in the interval 0 to 1.
Figs. 2a and 2b show the movement of the poles for solution I
found in ref. \cite{km1900}. As $\xi \to 0$, the decay
processes and corresponding mixing are switched off; the
positions of $K$-matrix poles
show the masses of the bare states: $f_0^{bare}(720 \pm 10)$,
$f_0^{bare}(1230 \pm 50)$,
$f_0^{bare}(1260 \pm 30)$,
$f_0^{bare}(1600 \pm 50)$ and $f_0^{bare}(1810 \pm 30)$.
In solution I, the gluonium state is associated with
$f_0^{bare}(1230 \pm 50)$ (it is $f_0^{bare}(1600 \pm 50)$
for solution II).
The case $\xi =1$ shows the
pole positions for the physical states $f_0(980)$, $f_0(1300)$,
$f_0(1500)$,
$f_0(1530{+90 \atop-250})$ and $f_0(1750)$.
In both solutions, according to \cite{n/d}, the broad resonance
$f_0(1530{+90 \atop-250})$ carries roughly $40\%$ - $50\%$ of the
gluonium component. This should not come as a surprise. Stationary
$q\bar q$ states are orthogonal to one another (though the loop
diagrams involving their decay cause some mixing), while the glueball
may mix freely with $q\bar q$ states; the rules of the $1/N$-expansion
\cite{1/n} tell us that $q\bar q$/gluonium mixing is not suppressed,
see \cite{n/d,ufn}. That is the reason for an accumulation of the
widths of the neighbouring $q\bar q$ states by the gluonium.

{\bf Broad resonance as a locking state.}
We now discuss the role of the broad resonance in the formation of the
confinement barrier. Let us consider as a guide the meson spectrum in
the standard quark model with decay processes switched off (Fig.\
3(a)). Here we have a set of stable levels. If the decay processes
are incorporated into highly excited states it would be naive to think
that the decay processes result only in broadening of levels. Due to
processes $bare\; state\to real\; mesons$, the resonances mix and one
of them may transform into the very broad state. This creates a trap
for the states with which it overlaps.

Fig.\ 3(a) shows the conventional potential model with a linear
potential rising indefinitely. That is unrealistic, since it does not
allow decays. In Fig.\ 3(b) we show instead a confinement barrier
through which states may decay, thus creating a broad locking state.
The broad resonance prevents decay of other states, which are left in
the small-$r$ region, the broad state plays the role of a dynamical
barrier. This is a familiar phenomenon: an absorption process acts as
a reflecting barrier. It means that comparatively narrow locked states
lie inside the confinement well, while the broad locking state appears
mainly outside the well. Experimental data confirm this idea.

Direct experimental evidence that the broad $0^{++}$ state is
associated with large $r$ is given by the GAMS data on $\pi ^- \pi ^+
(t) \to \pi ^0 \pi ^0$ \cite{gams}. The broad state is clearly visible
at $|t| < 0.2$ GeV$^2$, but disappears at large $|t|$, leaving peaks
related to narrower states clearly visible. This effect is
demonstrated in Fig.\ 4, which depicts $|A_{\pi \pi (t) \to
\pi\pi}(M)|^2$ found in \cite{km1900} for different $t$.

By fitting these data we determine the ratios of $t$
-distributions for narrow resonances and background. We find the
ratios:
\begin{eqnarray}
\frac{d\sigma_{\pi \to f_0(980)}(t) }{dt}&/&
\frac{d\sigma_{\pi \to f_0(1530{+90 \atop-250})}(t)}{dt}\ ,
\nonumber \\
\frac{d\sigma_{\pi \to f_0(1300)}(t) }{dt}&/&
\frac{d\sigma_{\pi \to f_0(1530{+90 \atop-250})}(t) }{dt}
\end{eqnarray}
as functions of $t$.
Assuming an universal $t$-exchange mechanism for all $f_0$ mesons,
these ratios are given by ratios of the transition form factors
squared:
\begin{eqnarray}
F^2_{\pi \to f_0(980)}(t)&/&F^2_{\pi\to f_0(1530{+90\atop-250})}(t)
\ , \nonumber \\
F^2_{\pi\to f_0(1300)}(t)&/&F^2_{\pi\to f_0(1530{+90\atop-250})}(t)
\end{eqnarray}

A method for calculating transition form factors at small and
moderate momentum transfers is given in \cite{ff}. Following it we
estimate mean radii squared of the $f_0$ -mesons using a simple
parametrisation of the wave functions in an exponential form
(exponential approximation works sufficiently well for wave
functions of mesons in the mass below 1500 MeV). Another
simplification: in the triangle diagram which is responsible for the
transition form factor at moderately small $|t|$, we approximate the
light cone variable energy squared as $s=\frac{m^2 +
k^2_{\perp}}{x(1-x)} \simeq 4(m^2 + k^2_{\perp})$ (at small $|t|$ the
region $x \simeq \frac{1}{2} $ gives the main contribution). Then
\begin{eqnarray}
&&\frac{d}{dt}
\ln[\frac{d\sigma_{\pi \to f_0}(t) }{dt}/
\frac{d\sigma_{\pi \to f_0(1530{+90 \atop-250})}(t)}{dt}]
\nonumber \\
=&&\frac{d}{dt} \ln[F^2_{\pi \to f_0}(t) /
F^2_{\pi \to f_0(1530{+90 \atop-250})}(t)]
\nonumber \\
=&&\frac{1}{3}
(R^2_{\pi \to f_0} - R^2_{\pi \to f_0(1530{+90 \atop-250})} )\ ,
\end{eqnarray}
where the transition radius
squared, $R^2_{\pi \to f_0}$, is
determined as
\begin{equation}
R_{\pi \to f_0}^2 =
\frac{2R_{\pi}^2 R_{f_0}^2 }{\frac{5}{3} R_{\pi}^2 + R_{f_0}^2 }\ ;
\end{equation}
$R_{\pi}^2$, $ R_{f_0}^2 $ are pion and $f_0$-meson radii squared.
Constituent quarks of the pion and $f_0$ meson are correspondingly in
$S$- and $P$-waves: that results in the factor $5/3$ in the
denominaror of Eq.\ (6).

A fit to data for the ratios of Eq.\ (3) gives:
\begin{equation}
R^2_{\pi \to f_0(980)} \simeq R^2_{\pi \to f_0(1300)}, \; \;
\mbox{or} \; \;
R^2_{ f_0(980)} \simeq R^2_{ f_0(1300)} ,
\end{equation}
and
\begin{equation}
R^2_{\pi \to f_0(1530{+90 \atop-250})} -
R^2_{\pi \to f_0(980)} \simeq (8 \pm 2) \mbox{GeV}^{-2} \; .
\end{equation}
The last equality defines the correlation between $R^2_{ f_0(1530{+90
\atop-250})} $ and $R^2_{f_0(980)}$, provided the mean pion radius
squared is fixed. Fig.\ 5 demonstrates this correlation for $R^2_\pi
= 0.41$ fm$^2$. It is clearly seen that $R^2_{ f_0(1530{+90
\atop-250})} $ is distinctly larger than $R^2_{f_0(980)}$ or
$R^2_{f_0(1300)}$: it means that a broad state at large $r$ can
definitely play the role of the locking state.

The same effect may take place for other exotic hadrons. By this we
mean glueballs and hybrids with different quantum numbers. One
example is now known for $J^P = 0^{-+} $ \cite{Zerominus}, where a
very broad $0^{-+}$ state around 1800-2100 MeV ($\Gamma \simeq 1$
GeV) has been identified in radiative $J/\Psi $ decays. Branching
ratios for this state to $\rho \rho$, $\omega \omega$, $K^*K^*$ and
$\phi \phi$ channels are in approximate agreement with
flavour-blindness, suggesting once again a strong glueball component.
The paper \cite{Zerominus} also advances arguments for a broad
$2^{++}$ resonance at 2000-2400 MeV; we have been involved in the
analyses of two sets of experimental data, to be published shortly,
providing evidence for this broad $2^{++}$ state.

{\bf Conclusion.}
In the deconfinement of quarks of an exited $q\bar q$-level, there are
two stages: \\
(i) An inevitable creation of new quark-antiquark pairs which result
in production of white hadrons. This stage is the subject
of QCD and is beyond our present discussion. \\
(ii) The outflow of the created white hadrons and their mixing results
in the production of a very broad state. The broad resonance locks
other $q\bar q$-levels into the small-$r$ region, thus playing the
role of a dynamical barrier; this is the reason for calling the broad
resonance a locking state.

The bare states are the subjects of the quark/gluon classification.
Exotic hadrons like glueballs and hybrids mix readily with $q\bar q$
states and are good candidates to generate locking states in all
waves.

Rich physics is hidden in the broad states, and an investigation of
them is an important and unavoidable step in understanding the
spectroscopy of highly exited states and their confinement.

{\bf Acknowledgements.}
We are grateful to Valeri Markushin and Victor Nikonov for
useful discussions and comments.
This work was supported by INTAS-RFBR grant 95-0267.

\onecolumn

\begin{figure}
\centerline{\epsfig{file=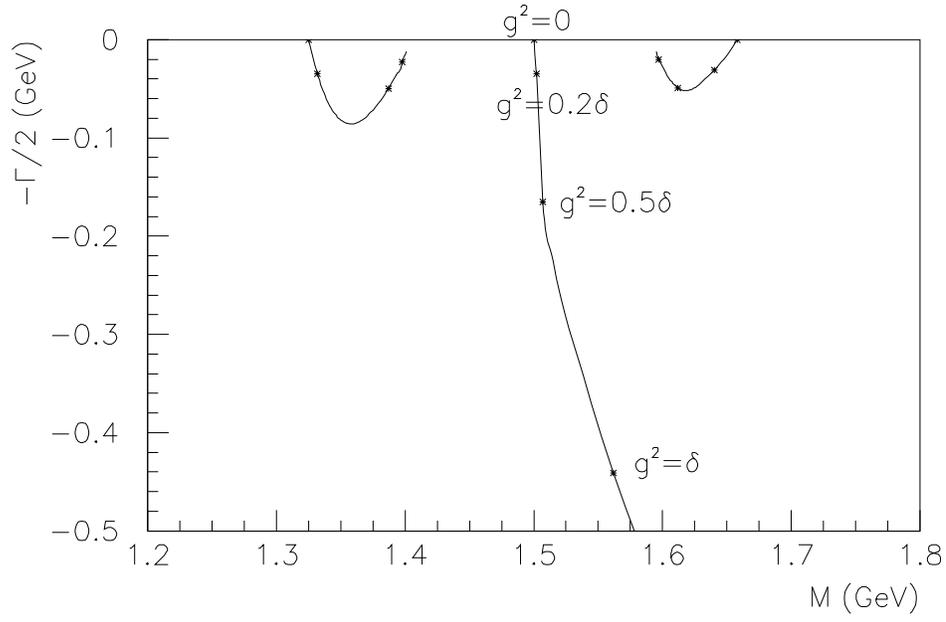,width=14cm}}
\vspace{-7cm}
\caption{Position of the poles of the amplitude of Eq.\ (1) in the
complex-$\sqrt s$ plane ($\sqrt s=M-i\Gamma/2$) with increase of
$g^2$; in this example $m=1.5$ GeV, $\delta=0.5$ GeV$^2$ and the phase
space factor is fixed: $\rho = 1$.}
\end{figure}

\begin{figure}
\centerline{\epsfig{file=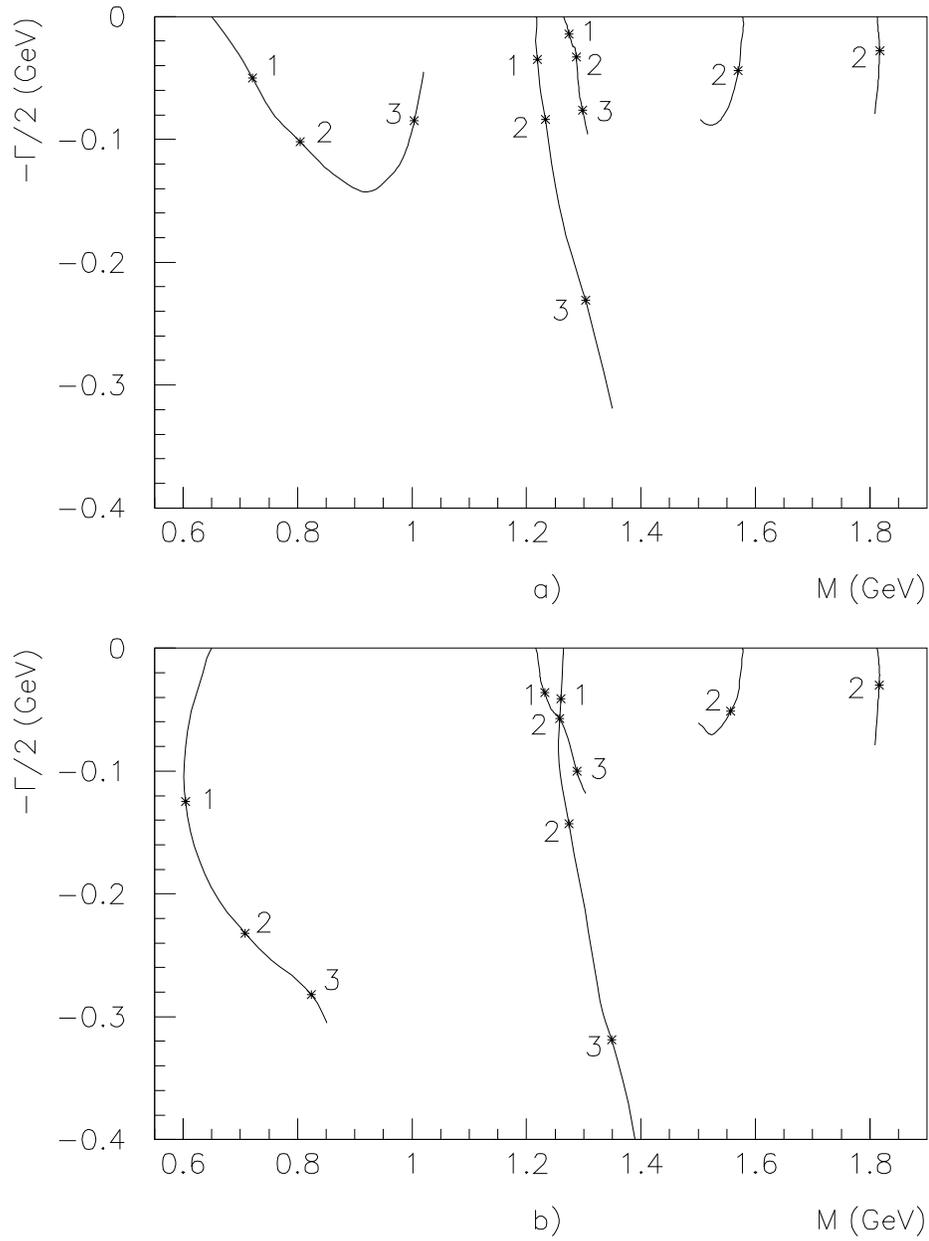,width=14cm}}
\caption{The pole positions of the $(00^{++})$-amplitude in the
complex-$\sqrt s$ plane ($\sqrt s=M-i\Gamma/2$), varying $\xi$: a)
the sheet under $\pi\pi$ and $\pi\pi\pi\pi$ cuts, b) the sheet under
$\pi\pi$, $\pi\pi\pi\pi$ and $K\bar K$ cuts. The crosses 1,2,3
indicate the values of $\xi$: 0.4, 0.6 and 0.9.}
\end{figure}

\begin{figure}
\centerline{\epsfig{file=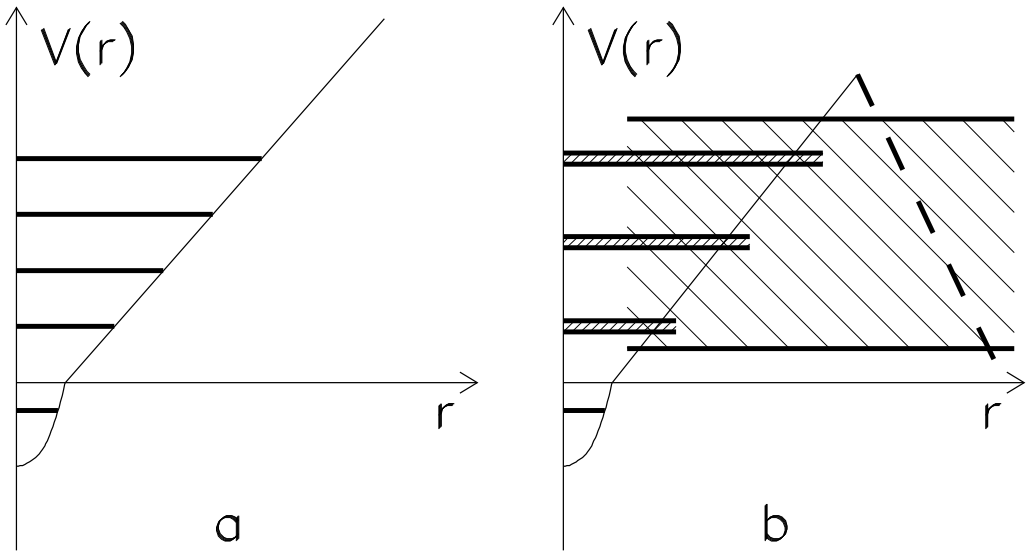,width=14cm}}
\caption{Conventional pictures for potential model levels
(a) without decay processes taken into account,
(b) with them.}
\end{figure}

\begin{figure}
\centerline{\epsfig{file=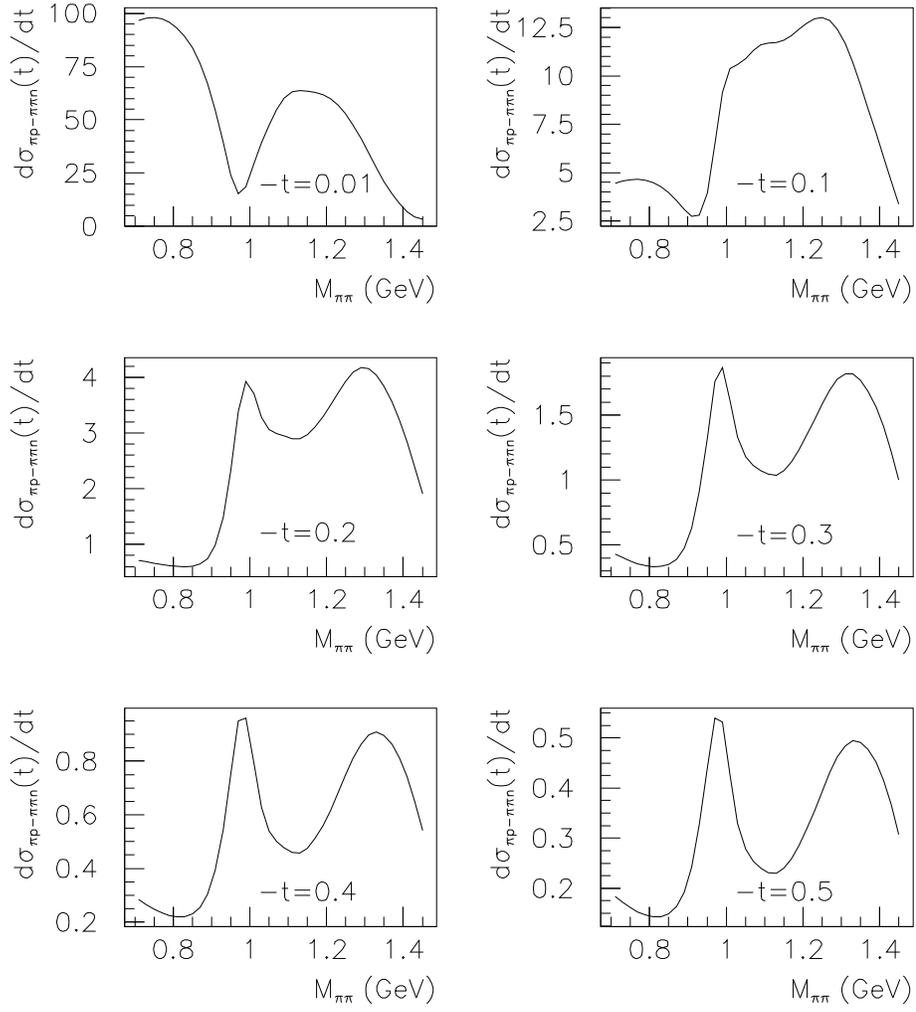,width=14cm}}
\caption{Mass spectra at different $t$ (GeV$^2$).}
\end{figure}

\begin{figure}
\centerline{\epsfig{file=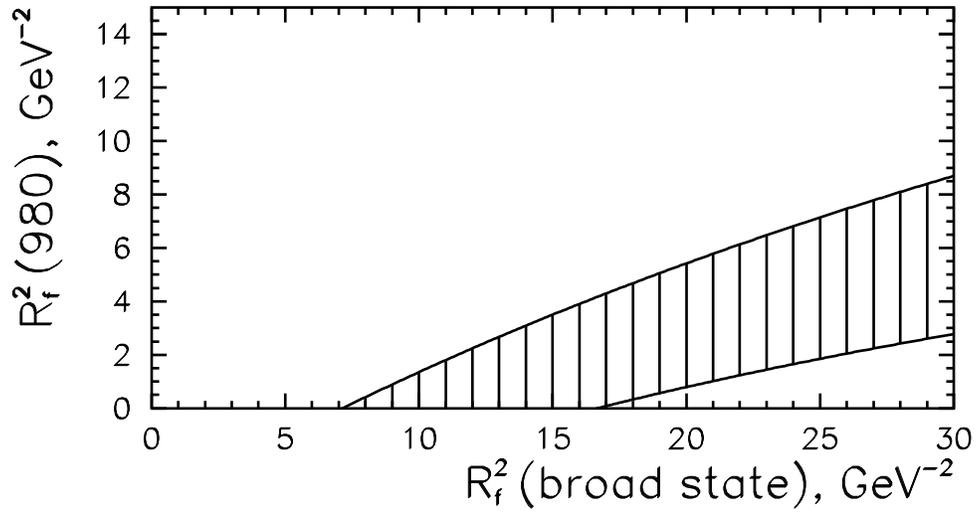,width=14cm}}
\caption{The correlation between radius squared of $f_0(980)$
and broad state $f_0(1530^{+90}_{-250})$.}
\end{figure}

\end{document}